\documentclass[preprint]{vgtc}               

\ifpdf
  \pdfoutput=1\relax                   
  \usepackage{graphicx}                
  \DeclareGraphicsExtensions{.pdf,.png,.jpg,.jpeg} 
\else
  \usepackage{graphicx}                
  \DeclareGraphicsExtensions{.eps}     
\fi%

\graphicspath{{figures/}{pictures/}{images/}{./}} 

\usepackage{microtype}                 
\PassOptionsToPackage{warn}{textcomp}  
\usepackage{textcomp}                  
\usepackage{mathptmx}                  
\usepackage{times}                     
\usepackage{cite}                      
\usepackage{tabu}                      
\usepackage{booktabs}                  

\usepackage{soul} 
\usepackage{balance}

\usepackage{xcolor,colortbl} 

\onlineid{-}

\vgtcinsertpkg

\title{Content Transfer Across Multiple Screens with Combined Eye-Gaze and Touch Interaction - A Replication Study}

\author{
Verena Biener \thanks{e-mail: verena.biener@hs-coburg.de}\\ %
      \scriptsize Coburg University of  \\
        \scriptsize Applied Sciences %
\and
Jens Grubert \thanks{e-mail: jens.grubert@hs-coburg.de}\\ %
        \scriptsize Coburg University of  \\
        \scriptsize Applied Sciences %
}

\abstract{In this paper, we describe the results of replicating one of our studies from two years ago \cite{Biener2020Breaking} which compares two techniques for transferring content across multiple screens in VR.
Results from the previous study have shown that a combined gaze and touch input can outperform a bimanual touch-only input in terms of task completion time, simulator sickness, task load and usability. Except for the simulator sickness, these findings could be validated by the replication.
The difference with regards to simulator sickness and variations in absolute scores of the other measures could be explained by a different set of user with less VR experience.%
} 


\teaser{
  \centering
  \includegraphics[width=\linewidth]{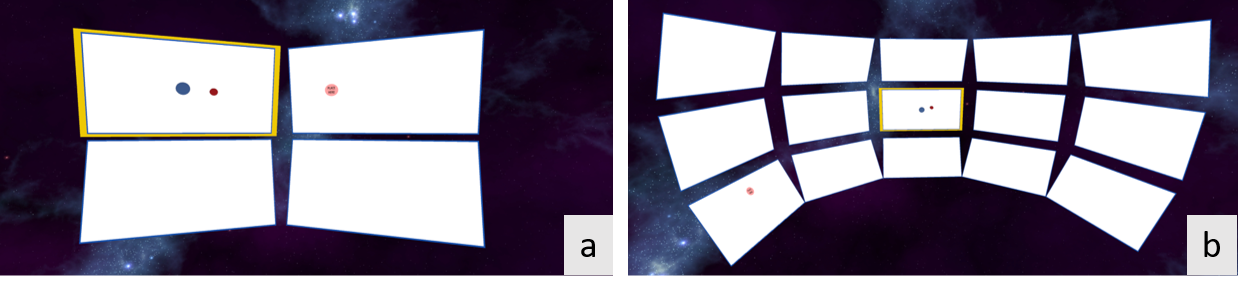}
  \caption{Arrangement of multiple screens in VR either with 4 (a) or 15 screens (15) centered around the users head.}
  \label{fig:teaser}
}

\begin{document}


\firstsection{Introduction}

\maketitle


In recent years researcher have become more aware of the replication crisis in empirical science. Disciplines like medicine or psychology are aware of this and encourage replication for quite some time \cite{moonesinghe2007most, carpenter2012psychology}.
Yet, it seems like there is less interest in replicating work in the field of HCI, as it is driven to publish novel results. 
However, some prior efforts were made to raise awareness in several RepliCHI workshops (e.g. \cite{wilson2013replichi, wilson2014replichi}) resulting in some replication attempts and experience reports \cite{wilson2014replichi}. 
Hornb{\ae}k et al. \cite{hornbaek2014once} looked at 891 HCI papers and found that only 3\% were attempting to replicate prior work and that a lot of papers try to downplay their replication part. The authors also mentioned that by just slightly changing the study design, many of then non-replication studies could have reinforced previous results.
In addition, Wacharamanotham et al. \cite{wacharamanotham2020transparency} analyzed how common it is to share research materials, as this can be an important contribution to facilitate replications. They found that only around 27–37\% of materials are shared.

The aim of this work is to replicate one of our previously conducted studies, to validate the findings and reflect upon the process of replication.
There are three different styles of replication defined by Hendrick in 1990 \cite{hendrick1990replications}. 
First, a conceptual replication uses a completely different approach  to evaluate the same hypothesis as in the original study. 
Second, in a partial replication only a part of the original study is changed, for example deleting or adding a variable.
Third, in a strict replication the experiment is replicated as exactly as possible.

We chose to do a strict replication of our previous work exploring, if combined touch and gaze interaction has benefits over touch-only interaction when interacting across multiple virtual screens \cite{Biener2020Breaking}.
Specifically, we replicate the content transfer task presented in the previous work.

\section{Interaction Techniques}
The previous work \cite{Biener2020Breaking} presented two interaction techniques for interaction with multiple displays in VR. 
Both techniques make use of a touchscreen for precisely controlling a cursor within one of the screens. In that case a one to one CD ratio is used to map the movement on the touchscreen to the virtual screen. 
Yet techniques differ in how the user can switch the active screen (indicated by the yellow frame in Figure \ref{fig:teaser}) and therefore moving the cursor to that screen.

\begin{figure}[t]
	\centering 
	\includegraphics[width=1\columnwidth]{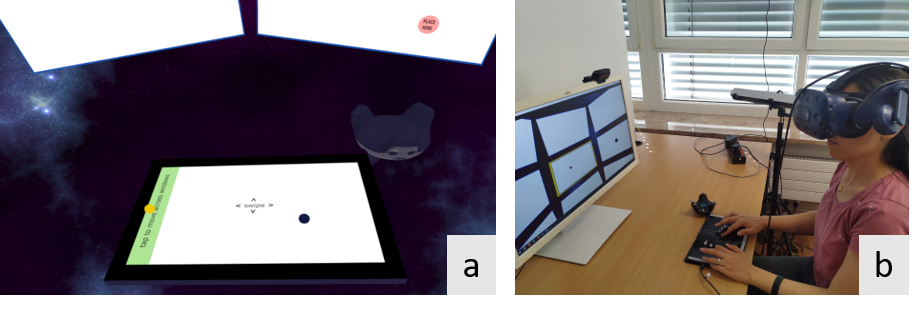}
	\caption{a) View on tablet when using the bimanual technique. Green area needs to be touched with the finger of the non-dominant hand (yellow sphere) to swipe through the screens using the finger of the dominant hand (blue sphere). b) Participant during the study.}
	\label{fig:setup}
\end{figure}

\paragraph{Bimanual:}
The bimanual technique uses the finger of the non-dominant hand on the border area of the touchscreen, with a width of 2 cm, as a mode switch (green area in Figure \ref{fig:setup}a). This means, while touching the border with the non-dominant hand, the dominant hand can be used to swipe through the screens and therefore moves the cursor to another screen. For every 2 cm that is swiped, the active screen will jump to the next one. This ratio was determined by Biener et al. \cite{Biener2020Breaking}.

\paragraph{Gaze-Based:}
In the gaze-based method the active screen is determined by gaze-direction. This means, the active screen is always the one the user is looking at. To facilitate looking at and selecting objects in the border area of a screen, a threshold of 5\% of the screen-size was added to delay changing the active screen.

\section{Study Design}
As in previous work \cite{Biener2020Breaking} the study was designed as a within-subjects study with two independent variables \textsc{Interaction Technique} and \textsc{Number of Screens}.
\textsc{Interaction Technique} was either \textsc{gaze-based} or \textsc{bimanual}.
\textsc{Number of Screens} was either \textsc{four} arranged in two rows (Figure \ref{fig:teaser}a) and two columns, or \textsc{fifteen} arranged in 3 rows and 5 columns (Figure \ref{fig:teaser}b).

The dependent variables were task load measured by NASA TLX (unweighted version) \cite{hart1988development}, usability as measured by the System Usability Scale (SUS) \cite{brooke1996sus}, simulator sickness (SSQ) \cite{kennedy1993simulator}, task completion time as the time between grabbing the object and releasing it on the target area and accuracy measured as the euclidean distance between the target area and the position where the object was placed. Additionally, users were also asked about their preferences.

\subsection{Task}
Depending on the condition, the participant is presented with either 4 or 15 screens, as presented in figure \ref{fig:teaser}. 
Upon beginning the task, a blue circular object will appear on one of the screens and a circular target area on another screen. It is the participants task to move the red cursor to the object using the touchscreen. Then, the participant should grab the object via long-press and move it to the target area where it can be released with another long-press.
Moving the cursor (and the object) across screens is done using the techniques previously described, depending on the condition.
Each new task started with the object on the screen where it was previously released. 
Exactly as in the previous study, the new target area was chosen pseudo-randomly making sure that different distances between initial position and target area are covered and both the object and the target area appeared on one of eight random positions on the screen.

\subsection{Apparatus}
As in the previous study \cite{Biener2020Breaking}, the experiment took place in a laboratory environment.
Also, the HTC Vive Pro eye was used in combination with a Huawei Media pad T5, with a screen diagonal of 10.1 inches and a 16:10 aspect ration and velcro on the border of the tablet supports the participants in finding the tablet boundaries.
During the study, the tablet was placed in front of the user on the table.
We used the exact same application as in the previous study, which was implemented in Unity.
The HMD was tracked with the Vive lighthouses and the fingers and the tablet were tracked using an Optitrack Trio system placed below the users head, to not interfere with the tracking of the HMD.
The size of the virtual screens was 24 inches placed at a distance of 90 cm around the participants head. The horizontal angle spanned by the screens was 190 degrees horizontally and 65 degrees vertically.

\subsection{Participants}
14 participants took part in this study (5 female, 9 male). Their mean age was 27.07 years ($sd=5.99$).
The mean height of the participants was 172.07 cm ($sd=10.74$).
Two people had no prior VR experience. 
Three people used VR once, four used it rarely, four sometimes and one often. No one reported to use it very frequently.
One participant never plays video games, six rarely do so, five play games sometimes, one often and one participant very frequently.
Eight participants wore glasses or contact lenses during the study.
Only one participant was left-handed, yet he uses his right hand for touch-input, so all participants used their right hand to control the cursor.
Two participants used their middle finger to touch and the other 12 used their index finger.
Only two participants also took part in the previous study.


\subsection{Procedure}
This replication study was conducted by the same person as in the original study.
First participants filled out a consent form and a demographic questionnaire.
Then the HTC Vive eye-tracking was calibrated.
After that the participants completed all four conditions in different orders, as determined by a balanced latin square (this resulted in 4 different orderings).
For each condition the participant could train the techniques in a short training phase with up to 10 tasks.
After that, they performed the actual condition with 32 tasks.
After completing each condition, participants filled out the system usability questionnaire (SUS), the simulator sickness questionnaire (SSQ) and the NASA task load index.
Upon completing all four conditions, a questionnaire about preferences was filled out and a short interview was conducted, asking participants about their preferences.
Participants were either taking part during work hours or they were compensated with 10 euros.

\section{Results}
As in the previous study, repeated measures analysis of variance (RM-ANOVA) with Holm-Bonferroni adjustments for multiple comparisons (initial significance level $\alpha=0.05$) was used to analyze the task completion time.
The subjective data obtained through questionnaires was analyzed using aligned rank transform before applying RM-ANOVA.

\subsection{Performance}
\begin{figure}[tb]
	\centering 
	\includegraphics[width=0.5\columnwidth]{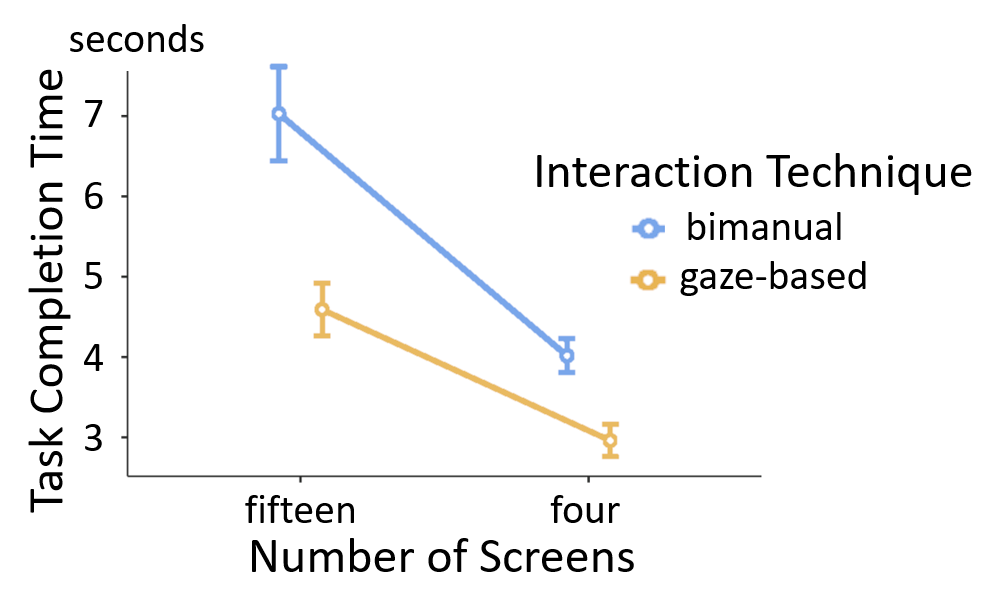}
	\vspace{-0.3cm}
	\caption{Task Completion Times of the four different conditions.}
	\label{fig:tct}
\end{figure}

The ANOVA results can be found in table \ref{tab:results}.
We could find significant main effects of the \textsc{Interaction Technique} on task completion time.
Specifically, \textsc{gaze-based} ($m=3.78s, sd=1.3$) resulted in a significantly lower task completion time than \textsc{bimanual} ($m=5.52s, sd=2.23$).
Also, the \textsc{Number of Screens} had a significant influence on task completion time in such a way that it was significantly shorter for \textsc{four} ($m=3.49s, sd=0.93$) than for \textsc{fifteen} ($m=5.81s, sd=2.14$).
In addition, the analysis also indicated interaction effects between \textsc{Interaction Technique} and \textsc{Number of Screens}. Looking at the graph in Figure \ref{fig:tct} it can be seen that for \textsc{four} screens the difference between \textsc{gaze-based} and \textsc{bimanual} is less prominent than for \textsc{fifteen} screens, yet, post-hoc tests still indicate it is significant ($p=0.001$).

To analyze the accuracy, we used the log-transform of the data as it was not normally distributed.
The ANOVA results indicated no significant influence of  \textsc{Interaction Technique} or \textsc{Number of Screens} on the accuracy.


\begin{table*}[t]
    \centering 
    \caption{RM-ANOVA results for Task Completion Time, Accuracy, Simulator Sickness, Task Load and System Usability. Cells with significant findings are marked grey. d$f_1$ = d$f_{effect}$ and d$f_2$ = d$f_{error}$.}
    \small
    \setlength{\tabcolsep}{5pt}
        
        \begin{tabular}{|c||c|c||c|c|c||c|c|c||c|c|c||c|c|c||c|c|c|}
            \hline 
            &\multicolumn{2}{|c|}{ }&\multicolumn{3}{|c|}{Task Completion Time} &\multicolumn{3}{|c|}{Accuracy} &\multicolumn{3}{|c|}{Simulator Sickness} &\multicolumn{3}{|c|}{Task Load} &\multicolumn{3}{|c|}{System Usability} \\
            \cline{2-18} 
            & d$f_{1}$ & d$f_{2}$ & F & p &  $\eta^2_p$  
            & F & p &  $\eta^2_p$ 
            & F & p &  $\eta^2_p$
            & F & p &  $\eta^2_p$
            & F & p &  $\eta^2_p$ \\ 
            \hline 
            Interface & $1$   & $13$  & $51.89$  &    \cellcolor{lightgray}$<0.001$ &  $0.8$    
            & $0.13$  &    $0.73$ &  $0.01$  
            & $0.53$  &    $0.48$ &  $0.04$  
            & $5.26$  &    \cellcolor{lightgray}$0.04$ &  $0.29$
            & $10.38$  &    \cellcolor{lightgray}$0.01$ &  $0.44$ \\  
            \hline 
            Number of Screens &  $1$  & $13$  & $37.13$  &  \cellcolor{lightgray}$<0.001$  & $0.74$     
            & $1.8$  &    $0.2$ &  $0.12$  
            & $3.48$  &    $0.08$ &  $0.21$  
            & $4.1$  &    $0.06$ &  $0.24$
            & $0.36$  &    $0.56$ &  $0.03$  \\  
            \hline
            Interface * Number of Screens & $1$   & $13$  & $5.14$  & \cellcolor{lightgray}$0.04$  & $.28$   
            & $0.72$  &    $0.41$ &  $0.05$ 
            & $2.39$  &    $0.15$ &  $0.16$ 
            & $0.07$  &    $0.80$ &  $0.04$
            & $0.62$  &    $0.45$ &  $0.05$  \\  
            \hline 
        \end{tabular}

    \label{tab:results}
\end{table*}

\subsection{Simulator Sickness, Workload, Usability}
No significant influence of \textsc{Interaction Technique} or \textsc{Number of Screens} on simulator sickness was found. Yet, \textsc{bimanual} ($m=22.31, sd=19.49$) resulted in slightly higher measures than \textsc{gaze-based} ($m=20.84, sd=21.04$).

Regarding the overall taskload, the results indicate that \textsc{Interaction Technique} had a significant influence.
Specifically, \textsc{gaze-based} ($m=26.46, sd=12.12$) resulted in a significantly lower taskload than \textsc{bimanual} ($m=33.72, sd=16.74$).
In contrast to the original study, no significant effect could be detected on the performance aspect.

The results also indicate that \textsc{Interaction Technique} has a significant influence on usability, such that \textsc{gaze-based} ($m=87.5, sd=10.8$) resulted in a significantly higher usability than \textsc{bimanual} ($m=78.66, sd=14.84$).


\subsection{Preference, Comments, Observations}
When asked which method they prefer, all but one participant answered \textsc{gaze-based}.
This participant also felt like \textsc{bimanual} was faster and more accurate.
All other participants said that \textsc{gaze-based} was faster and more accurate, except one person found \textsc{bimanual} more accurate in the condition with \textsc{four} screens and one person found it more accurate in the condition with \textsc{fifteen} screens.
Participants preferring \textsc{gaze-based} said it was easier (5 participants), faster (5 participants), more intuitive (1 participant), the was no need to coordinate hands (2 participants) and it resulted in generally less physical movement (4 participants).
The participant who preferred \textsc{bimanual} said that he found it more precise and it was impractical for him that the screen directly changed by just looking around.




\section{Discussion: Comparison to Previous Results}
When comparing the results of the replicated study to the original one, it can be seen that they are very similar.
Both times, a significant main effect of \textsc{Interaction Technique} and \textsc{Number of Screens} on the task completion time could be found, such that \textsc{gaze-based} is significantly faster than \textsc{bimanual} and \textsc{four} screens are significantly faster than \textsc{fifteen}.
Also the absolute task completion times are very similar in both studies (original study: \textsc{gaze-based}$=3.7s$, \textsc{bimanual}$=5.49s$; replication study: \textsc{gaze-based}$=3.78s$, \textsc{bimanual}$=5.52s$). 
Yet the replication study also revealed an interaction effect between \textsc{Interaction Technique} and \textsc{Number of Screens} indicating that for \textsc{four} screens the difference between \textsc{gaze-based} and \textsc{bimanual} is less prominent, yet still significant.
Both the original and the replication study did not reveal any effects on accuracy.

Regarding simulator sickness, the original study found a significant difference between \textsc{gaze-based} and \textsc{bimanual} interaction. This effect could not be seen this time, but still \textsc{gaze-based} resulted in a slightly better score than \textsc{bimanual}. 
However, simulator sickness is generally higher (55\%) in the replication study ($m=21.58$) than in the original study ($m=13.89$).
This could be explained by the experience of the users \cite{chang2020virtual}. In the previous study all participants had prior VR experience, while in the replication two had none. And while 6 participants in the previous study used VR often or very frequently, only 1 participant from the replication study uses VR often.

Both studies indicate a higher taskload for \textsc{bimanual}.   
The absolute values were generally slightly higher (16\%) in the replication study ($m=30.09$) than in the original study ($m=25.94$). It can again be speculated that this could be due to less experienced participants.

The results for usability were also the same in both studies, with \textsc{gaze-based} being considered significantly more usable than \textsc{bimanual}. 
The absolute scores for \textsc{gaze-based} are very similar between the original ($m=88.75$) and the replication ($m=87.5$). However, the \textsc{bimanual} technique was considered 8\% more usable in the replication ($m=78.66$) compared to the original study ($m=72.59$).

Comparing the effect sizes ($\eta^2_p$) shows that they were slightly higher for the data of the original study, especially for the subjective data from questionnaires. For example, the effect size of the main effect of \textsc{interface} on usability was previously $\eta^2_p=0.75$ while it was $\eta^2_p=0.44$ in the replication. And the effect size of the main effect of \textsc{interface} on task load was previously $\eta^2_p=0.7$ ans is $\eta^2_p=0.29$ in the replication. In addition, also the p-values indicated a higher significance level. 
It was $p<0.001$ for the effect of \textsc{interface} on usability in the original study and $p<0.01$ in the replication. And similarly it was $p<0.001$ for the effect of \textsc{interface} on task load previously and in the replication it was $p=0.04$.
However, this difference is not reflected in the performance data of task completion time, suggesting a distinct difference between the two groups of participants in how they perceive the system and their own performance.

Finally, in both studies there was only one participant who preferred the \textsc{bimanual} technique. The main reason for preferring \textsc{gaze-based} in the original study was that it is faster. This was also one of the main reasons this time, besides it being easier.

Another important aspect that we noticed when replicating the study setup was, how important it is to have clear instructions of the application and how to use it. Especially in a case like this where multiple systems need to work together, such as the tablet sending the touch data, the Optitrack system providing tracking data of the fingers and tablet and the Vive tracking system. 
A good documentation would be even more crucial when making code publicly available.






\section{Conclusion}
In this work, we replicated one of our previous studies comparing touch-only input to a multimodal approach combining touch with eye-gaze tracking for interacting with multiple displays in VR.
The general findings could be validated. Both the original and the replicated study indicate that the approach combining eye-gaze and touch results in faster task completion times, lower task load and higher usability ratings. However, previously found differences regarding simulator sickness could not be confirmed.
Yet, when looking at subjective ratings of simulator sickness and task load, it can be seen that the replication study resulted in generally higher scores. This could possibly be explained by the different set of participants with very apparent differences in VR experience.


\bibliographystyle{abbrv-doi}

\bibliography{template}
\balance
\end{document}